\documentclass{article}

\usepackage[final,nonatbib]{neurips_2023}

\usepackage[utf8]{inputenc} 
\usepackage[T1]{fontenc}    
\usepackage{hyperref}       
\usepackage{url}            
\usepackage{booktabs}       
\usepackage{amsfonts}       
\usepackage{nicefrac}       
\usepackage{microtype}      
\usepackage{xcolor}         
\usepackage[pdftex]{graphicx}
\usepackage{amsmath}
\usepackage{multirow}
\usepackage{listings}
\newcommand{\eg}{\textit{e.g., }}
\newcommand{\cf}{\textit{cf. }}

\newcommand{\ie}{\textit{i.e., }}

\newcommand{\sysws}{\textsc{BioSpark\ }}
\newcommand{\sys}{\textsc{BioSpark}}

\newcommand{\code}[1]{\texttt{\small{#1}}}

\title{\sys: An End-to-End Generative System for Biological-Analogical Inspirations and Ideation}

\author{
  Hyeonsu B. Kang\(^{\dagger}\), 
  David Chuan-En Lin\(^{\dagger}\),
  Nikolas Martelaro\(^{\dagger}\), 
  Aniket Kittur\(^{\dagger}\) \\
  \textbf{Yan-Ying Chen}\(^{\ddagger}\),
  \textbf{Matthew K. Hong}\(^{\ddagger}\) \\[1ex] 
  \(^{\dagger}\)Human-Computer Interaction Institute, Carnegie Mellon University, Pittsburgh, PA 15213\\[.5ex]
  \(^{\ddagger}\)Toyota Research Institute, Los Altos, CA 94022\\[1ex]
  $^{\dagger}$\fontsize{8}{8}{\texttt{\{hyeonsuk,chuanenl,nikmart,nkittur\}@cs.cmu.edu}}\ \ 
  $^{\ddagger}$\fontsize{8}{8}{\texttt{\{yan-ying.chen,matt.hong\}@tri.global}}
}

\begin{document}

\maketitle
\begin{abstract}
Nature is often used to inspire solutions for complex engineering problems, but achieving its full potential is challenging due to difficulties in discovering relevant analogies and synthesizing from them.
Here, we present an end-to-end system, \sys, that generates biological-analogical mechanisms and provides an interactive interface to comprehend and synthesize from them.
\sys{} pipeline starts with a small seed set of mechanisms and expands it using an iteratively constructed taxonomic hierarchies, overcoming data sparsity in manual expert curation and limited conceptual diversity in automated analogy generation via LLMs.
The interface helps designers with recognizing and understanding relevant analogs to design problems using four main interaction features.
We evaluate the biological-analogical mechanism generation pipeline and showcase the value of \sysws through case studies.
We end with discussion and implications for future work in this area.
\end{abstract}

\begin{figure}[h]
    \includegraphics[width=\linewidth]{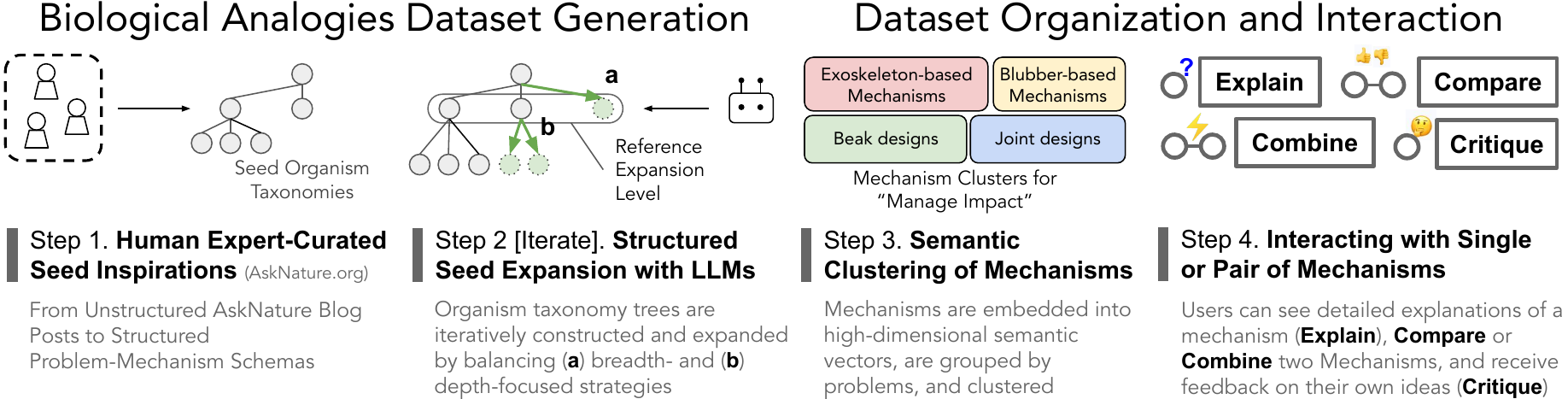}
    \vspace{-1.5em}
    \caption{\sys{} consists of a \emph{Dataset Generation Pipeline} and an \emph{Organization and Interaction} interface. Dataset construction starts from a small set of seed mechanisms on AskNature (\textbf{Step 1}), expanded by iteratively constructing taxonomic hierarchies of organisms and diversifying using \emph{breadth}- or \emph{depth}-focused knowledge prompts (\textbf{Step 2}). Dataset is organized by semantic clusters (\textbf{Step 3}) and provided with interactive features that facilitate user learning and synthesis (\textbf{Step 4}).}
    \label{fig:main}
    \vspace{-1.5em}
\end{figure}
\section{Introduction}
Nature provides a valuable source of inspirations~\cite{benyus1997biomimicry,BioTRIZ,shu2011biologically} for novel design solutions to challenging engineering problems~\cite{vandevenne2016enhancing,shu2011biologically}.
A central cognitive mechanism underpinning this transfer of domains is \textit{analogical processing}~\cite{gentnerAnalogyCreativityWorks1997,oppenheimerAnalogyScience1956,gickholyoak1980}, which involves {discovering} and {recognizing} deep structural similarity between source and target analogs.
To this end, recent work has shown that analogs with a similar purpose (\ie `\emph{what problem it solves}'), but diverse mechanisms (\ie `\emph{how it solves it}'), facilitate creative new ideas~\cite{analogy_search_engine,hope_kdd17}.

Yet, attaining the full potential of biological-analogical inspirations in engineering and design domains has proven difficult.
Many bio-inspired designs are biased toward well-known organisms~\cite{broeckhoven2022escaping} and prior work has shown it can take on average three person-hours to sift through vast, fragmented information sources such as Google Scholar, Wikipedia, AskNature.org, or the open Web to find a single relevant article for a design problem~\cite{vattam2011foraging,vattam2013seeking}.
Human-expert-curated databases such as AskNature~\cite{deldin2013asknature}, DANE~\cite{DANE}, SAPPhIRE~\cite{SAPPhIRE}, BioTRIZ~\cite{BioTRIZ}, or rule-based databases (\eg \cite{cheong2014retrieving}) can streamline searches, but the effort-intensive process of manual expert curation often results in data scarcity, limiting the diversity of organisms and mechanisms available for making analogies.
Automatic generation of analogies with LLMs could create larger datasets for bio-inspired design, but they still have limited diversity on abstract concepts, even when tuning parameters to increase diversity such as logic suppression and increased sampling temperature~\cite{chung_2023_llm_diversity}.
Knowledge-augmented prompting provides an alternative design space to token-level manipulation, as explored in~\cite{baek_2023_knowledge_augmented} for improving the factuality of Q\&A tasks.
We explore this design space for a different purpose of constructing a diverse biological mechanisms dataset, with a modular and structured knowledge augmentation approach.

Beyond sourcing biological analogies, it can be challenging for people to understand and recognize them when presented. 
People's cognitive biases towards surface similarity and working memory limitations can impede the mapping process between biological and engineering systems~\cite{gentner1985analogical,ross1987like,gentner1983structure}, while the need for situating found (cross-domain) analogies within the constraints and scope of the original design problem adds additional complexity for designers developing their own ideas (\cf~\cite{KANG2022AUGMENTING_NAACL,analogy_mining_design_needs_chi18_gilon}).
To mitigate these challenges, we construct a scalable analogy pipeline that balances the cost of expert-curation and the limitations of na\"{i}ve LLM-based generation, and develop an interactive system that aids designers in recognizing and understanding the relevance and adaptability of analogs for a chosen design problem.
\section{Usage Scenario} \label{section:usage_scenario}
Consider an automobile designer who is working on a new bike rack design for sedans.
The designer identifies the core challenge of designing a secure bike rack as turbulence management, caused by variable wind resistance due to changing vehicle speeds and additional factors from inclement weather conditions, such as precipitation.
When the designer clicks on the `Manage Turbulence' problem as a query (Fig.~\ref{fig:case_study_1}, top right), \sys{} displays mechanism clusters related to the problem.
She quickly identifies particularly interesting and idea-provoking mechanisms such as ``Adapative shape-shifting mechanism in intertidal microalgae'' and ``Friction-based attachment mechanism of parasitic copecods,'' two quite different mechanisms, yet commonly observed in organisms that live underwater.
When exposed to turbulence cues, shape-shifting microalgae adaptively reconfigure their surface shapes to reduce resistance from strong currents.
This provides her a design idea for adaptively ``lying down'' bike racks on the roof of the car that reduce air drag, and a ``tail basement'' bike rack attachment to the back of the car that joins the rotational axis of the front wheel of the bikes and adjusts during turbulence by flexibly (counter-)swinging bikes, mimicking bird tail motions.
She clicks on the `Combine' tab with the two mechanisms selected to experiment with design ideas.
The result describes an approach that simply combines the friction-based mechanism to the shape-shifting mechanism.
This gives her an additional idea for further securing the racks, by incorporating designs inspired from the adhesion mechanism of parasitic copecods, such as microridges and channels for increasing friction upon contact.
\section{System Description}
\begin{figure}[h]
    \vspace{-1em}
    \includegraphics[width=\linewidth]{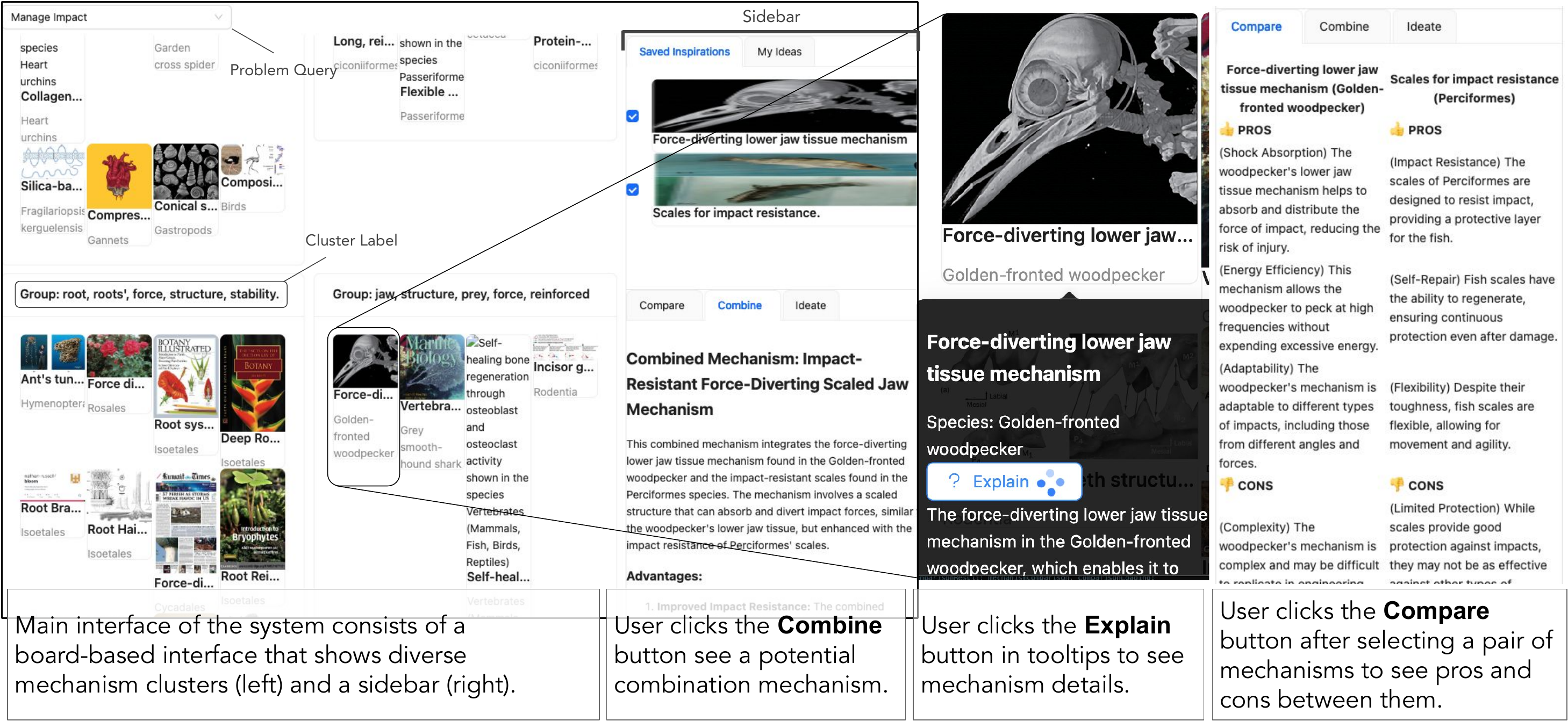}
    \vspace{-1.5em}
    \caption{\sys{} interface and a subset of available interaction affordances.}
    \label{fig:interface}
    \vspace{-1.5em}
\end{figure}
\subsection{Structuring AskNature blog posts' text into a seed set of problem-mechanism schemas} \label{appendix:schemas_from_asknature}
To source a set of diverse, high-quality biological mechanisms for a given problem, \sys{} starts from a seed set of expert-curated biological mechanisms on AskNature (Fig.~\ref{fig:main}, Step 1). 
AskNature.org provides a curated list of organisms with detailed descriptions of their unique strategies to functional problems (\eg `Manage Impact', `Modify Speed').
The organisms and strategies can be grouped by function and viewed as a list.
To curate a seed set of high-quality mechanisms, we first choose a functional problem $p$ predicted to be highly relevant to automobile designers, excluding irrelevant functions such as `Adapt Behaviors', `Adapt Genotype', `Coevolve', `Maintain Community'
We access the sub-list of organisms $o \in O$ and strategies posted to $p$ on AskNature's group-by-function page by parsing the \code{HTML} code using the \code{BeautifulSoup} package on \code{Python}. We then access the blog post for each organism-strategy page using the parsed URL and parse the returned \code{HTML} page to get the title, description, and references (if available).

At this stage, the returned unstructured text is yet to contain a succinct mechanism description.
Furthermore, we found that some blog posts do not contain any body text despite having a title and are accessible via the URL.
Some of these missing blog posts indicated that they are in maintenance and/or planned to be updated.
To structure the raw blog post text $\text{AskNature}_{(o, p)}$, we prompt GPT4~\cite{openai2023gpt4} to succinctly describe (\ie using 12 words or less) the core mechanism (\ie excluding the qualities or effects, and focusing on mechanisms with engineering design implications), given $(o, p)$ (if blog post text is missing) or $(o, p, \text{AskNature}_{(o, p)})$.
The returned mechanism description $m$ along with the function description makes up the problem-mechanism schema for each organism: $\{o \in O|(p, m, o)\}$.

\subsection{Iteratively constructing taxonomic trees with organisms represented in found mechanisms and performing structured expansion of a tree} \label{appendix:iterative_structured_expansion}
Using each schema as a seed, we iteratively prompt GPT4 to find relevant mechanisms for the given mechanism and problem, using an even mixture of breadth- and depth-focused expansion strategies (Fig.~\ref{fig:main}, Step 2).
To enable structured diversification of organisms and their mechanisms beyond prior work that relied on token-level manipulation or na\"{i}vely prompting LLMs, we guide LLMs \emph{how} and \emph{where} to expand by leveraging organism taxonomic hierarchies.
At each iteration of expansion (Fig.~\ref{fig:main}, Step 2), we aggregate the organisms represented in found mechanisms up to that point, and construct a taxonomic tree featuring seven levels of hierarchy on Tree of Life: \{\code{domain, kingdom, phylum, class, order, family, genus, species}\}, where \code{domain} representing the highest level and \code{species} representing the lowest level on the hierarchy.

Given this tree, we aim to identify sparsely populated branches for expansion.
We cut the tree at a given reference expansion level (\eg \code{class}), and sort the taxonomic ranks (nodes) on that level by the number of its immediate children nodes\footnote{alternatively, the entire size of the subtree, rather than immediate children, could be used for sorting}, in an increasing order. 
For performance, we batch 10 prompts to send to GPT4 for expansion.
For half of the prompts, we instruct \textbf{breadth-first expansion} which asks GPT4 to first identify \emph{sibling} nodes at the given reference taxon level and existing nodes (up to 50 most populated nodes).
For example, the prompt asks ``come up with a few biological \code{classes} not in \{\code{...names of excluded classes...}\}''.
The breadth-first expansion prompt then instructs GPT4 to repeat the following: 1) Choose one taxon from the list it came up with; 2) Succinctly describe (\ie using 14 words or less) new mechanisms $m$ related to a problem $p$.
For the rest of the prompts, we instruct \textbf{depth-first expansion} which asks GPT4 to first identify a new \emph{children} node at the given reference taxon level and existing children nodes (up to 50 randomly sampled children).
For example, the prompt asks ``come up with a few biological \code{families} in \code{order} \code{araneae} that are not any of \{\code{araneidae}, ...\}''.
The depth-first expansion prompt then instructs GPT4 to repeat a similar procedure as breath-first expansion.
In the prototype system, we run 10 batches for expansion to construct dataset of mechanisms for each problem.

The returned list of mechanisms and organisms text are then fed into the second GPT4 prompt for structuring them into a list of \{\code{mechanism, organism}\} dictionaries.
Finally, using each organism name, we prompt GPT3.5-turbo to retrieve the seven-level taxonomic hierarchy, based on our model evaluation result showing its high accuracy (\S\ref{subsection:accuracy_taxonomy_construction}).

\subsection{Semantic Clustering of Mechanisms} \label{appendix:semantic_clustering}
The mechanism descriptions are clustered in semantically related groups for designers' efficient viewing (Fig.~\ref{fig:main}, Step 3).
We first embed the mechanism descriptions for a given problem into high-dimensional vectors using the \code{text-embedding-ada-002} end-point of the OpenAI API.
We group the embeddings into $k=20$ clusters using k-Means clustering.
Each group label is generated by chunking the mechanism descriptions in each cluster into lower-cased words, excluding stopwords.
We count the frequency of each word and take the top 5 most frequent words as a description of each cluster.

\subsection{Representative Mechanism Image Curation} \label{appendix:image_search}
To aid designers' visual understanding of and pique curiosity for biological-analogical mechanisms, we retrieve representative images for corresponding textual mechanism descriptions.
We use Google Custom Search\footnote{\url{https://developers.google.com/custom-search/v1/overview}} with queries as ``\code{[organism name]:[mechanism description]}'' and the file type set to images and the safe search mode enabled.
We choose the first place result of Custom Search as the visual representation of each mechanism.

\subsection{Interacting with Mechanism Inspirations: Explain, Compare, Combine, and Critique} \label{appendix:interaction_features}
To facilitate designers' understanding and synthesis of mechanism inspirations, we develop several interaction features available on the interface (Fig.~\ref{fig:interface}).
The \textbf{Explain} button is located in tooltips that pop up when the user places the mouse over on a mechanism card in the board UI (Fig.~\ref{fig:interface}, first panel).
When the user clicks on the button, \sys{} sends a prompt to GPT4 requesting elaboration of the interacted mechanism and the organism in the context of the chosen engineering design problem.
The \textbf{Compare} tab is located in the control bar of the sidebar of the interface.
To use this, users need to first click on (at least) two mechanism cards from the left, saving them to the `saved inspirations' panel at the top of the sidebar.
There, users can check any two of the saved mechanisms they wish to compare.
\sys{} sends a prompt to GPT4 when the user clicks on the tab, requesting comparison of pros and cons between the two mechanisms in the context of the chosen engineering problem.
The result is formatted into a markdown table, with each mechanism as the header followed by pros and cons rows detailing each point.
The \textbf{Combine} tab is also located in the control bar of the sidebar in the interface.
Similarly with Compare, users can check two saved mechanisms they wish to see combined.
\sys{} sends a prompt to GPT4 then requesting elaboration of a mechanism that combines the two selected mechanisms, and explain its potential advantages in the context of the chosen engineering problem.
The result is also formatted into a markdown page using section title and headers for demarcating the content.
Finally, the \textbf{Critique} button is located inside the Ideate tab.
Upon clicking the Ideate tab, users can type in their own idea in the rich text editor in the opened tab, and optionally click on the button below to receive constructive feedback on it.
\sys{} sends a prompt to GPT4 with the content of the text editor describing the idea, and requests additional revision that may improve the quality, such as anticipated failure modes and potential improvements.
\section{Evaluation}
\subsection{Accuracy of LLM-based Taxonomy Construction} \label{subsection:accuracy_taxonomy_construction}
The main process in our diversification strategy is iterative construction of taxonomic trees at each stage of expansion with a set of problem-mechanism schemas and corresponding organisms $\{o \in O|(p, m, o)\}$ curated (in case of AskNature seeds) or generated up to that point.
To construct the trees, the taxonomic hierarchy of each organism needs to be known.
Here, we restrict our tree construction to seven levels of depth, ranging from the highest to lowest levels: \code{domain, kingdom, phylum, class, order, family, genus, species}. 
These levels provide considerable branch-switching opportunities for diversification, through significant changes in the number of members between levels and within each level of the hierarchy.
For example, while the highest level \code{domain} consists of three members, Bacteria, Archaea, and Eukarya, there are estimated 8.7M species in the world~\cite{sweetlove2011}. 
The next level on the hierarchy, \code{Genus}, has an estimated number of 310K members~\cite{rees2020all}, while the number in the subsequent level, \code{families}, is estimated at 8K~\cite{mora2011many} in 2011.
The number of known species for each node on the hierarchy also changes considerably, further contributing to the diversification opportunities.
For example while most non-avian reptile genera have only 1 species each, insect genera such as \textit{Lasioglossum} and \textit{Andrena} have over 1,000 species each, while the flowering plant genus, \textit{Astragalus}, contains over 3,000 known species~\cite{wiki:genus}.

Our initial exploration of suitable approaches to retrieve organism taxonomies involved using available resources such as the Global Biodiversity Information Facility API\footnote{\url{https://www.gbif.org/developer/species}}, Catalogue of Life~\cite{banki2023}, or the Encyclopedia of Life~\cite{eol}, where canonical species names were retrieved from the Darwin Core List of Terms\footnote{\url{https://dwc.tdwg.org/list/\#dwc\_Organism}} for corresponding organisms in problem-mechanism schemas.
However, the limited coverage, data consistency, and API availability of these tools prevented their adoption.
On the other hand, Wikipedia provides scientific classification for some of the organism articles (for example in the Pomelo article\footnote{\url{https://en.wikipedia.org/wiki/Pomelo}}, taxonomic names for \code{Kingdom, Clade, Order, Family, Genus,} and \code{Species} are available in the `biota' information box that appears on the right-hand side of the page).
However, this data was not readily available for scalable generation.

\subsubsection{Procedure}
LLMs may provide an alternative solution to the limitations of existing approaches for retrieving the taxonomic hierarchy for a given organism name.
To test this idea, we curated 90 gold taxonomies using Wikipedia that have complete information in the `biota' scientific classification info box.
For each organism, we prompted LLMs with each organism name zero-shot using the chat completions API endpoint\footnote{\url{https://api.openai.com/v1/chat/completions}} using each model key (see Appendix~\ref{appendix:taxonomy_prompt} for the prompt used).
Once the hierarchy data is generated, we lower-cased the rank names for consistency.
The complete list of organism names used for generation can be found in Appendix~\ref{appendix:complete_list_of_organisms}.

\begin{table*}[t!]
    \centering
    \begin{tabular}{p{1.8cm} p{1.3cm} p{1.3cm} p{1.3cm} p{1.3cm} p{1.3cm} p{1.2cm} p{1.2cm}}
    \toprule
    \textbf{Model} & \textsc{Domain} & \textsc{Kingdom} & \textsc{Phylum} & \textsc{Class} & \textsc{Order} & \textsc{Family} & \textsc{Genus} \\
    \midrule
    \multirow{2}{*}{\code{GPT4}} & 100\% (90/90) & 100\% (90/90) & 100\% (90/90) & 100\% (90/90) & 96.7\% (87/90) & 94.4\% (85/90) & 98.9\% (89/90) \\
    \multirow{2}{*}{\code{GPT3.5-turbo}} & 100\% (90/90) & 100\% (90/90) & 100\% (90/90) & 100\% (90/90) & 95.6\% (86/90) & 95.6\% (86/90) & 93.3\% (84/90) \\
    \bottomrule
    \end{tabular}
    \caption{The accuracy of zero-shot taxonomy generation using only the organism name.}
    \vspace{-2em}
    \label{table:taxonomy_generation_accuracy}
\end{table*}
\subsubsection{GPT4's Accuracy}
We find that GPT4's zero-shot taxonomy generation accuracy to range between 94.4\% and 100\% (Table~\ref{table:taxonomy_generation_accuracy}).
The lowest accuracy was observed in the \code{family} taxonomy, followed by \code{order} (96.7\%) and \code{genus} (98.9\%).

\subsubsection{System Optimization: GPT3.5-turbo's Accuracy}
We find that GPT3.5-turbo has comparable accuracy levels with GPT4 in zero-shot taxonomy generation. 
The highest misaglignment occurred in \code{genus}, with a 6.67\% error rate (equivalent to 6 out of 90).
Appendix~\ref{appendix:taxonomy_generation_error_analysis} provides a further qualitative error analysis of models' comparative performance.
Based on these results, we opted for the more efficient GPT3.5-turbo model in our pipeline.
We leave further exploration of the capabilities of smaller, fine-tuned base LLMs, with implications for LLM cascade\footnote{LLM cascade refers to a system design approach that adaptively chooses optimal LLM APIs for a given query. Smaller, task-specific LLMs are regarded as optimal when they exhibit higher or similar levels of performance compared to models that are orders of magnitude larger~\cite{dohan2022language}, with all else equal.}, to future work.

\subsection{Increase in Organism Diversity}
In order to evaluate the effectiveness of diversification through our expansion strategies from iteratively constructed taxonomic trees, we investigated how organism diversity changes upon a series of mechanism generation.
\begin{figure}[h]
    \includegraphics[width=\linewidth]{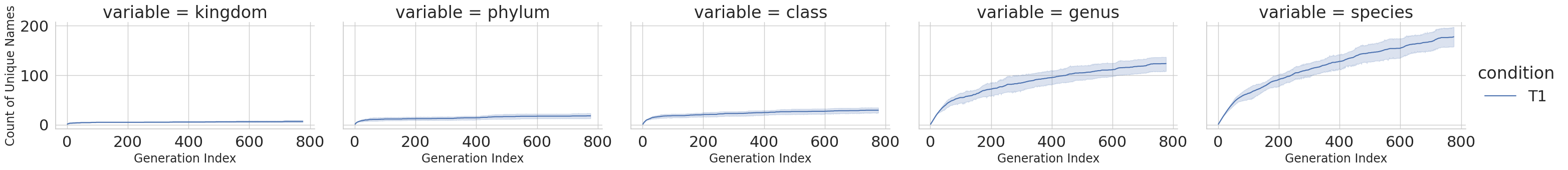}
    \vspace{-2em}
    \caption{Organism diversity, measured by the number of unique names among the generation, increases monotonically as generation continues, while the number of mechanisms generated per organism also increases, as evidenced by the decreasing slope. This suggests mechanism concentration over each organism increases over time.}
    \label{fig:organism_diversity}
\end{figure}
\subsubsection{Procedure}
We generate mechanisms and corresponding species for five problems closely related to automobile design: `managing impact', `managing tension', `managing compression', `managing turbulence', `modifying speed'.
We index the species at the time of its appearance in the corresponding mechanism generation.
Hence, the index corresponds to when a new mechanism was generated via our pipeline.
At each generation index, we count the unique number of names that are generated up to that point, for each taxonomic rank, and average the numbers across the five problems.

\subsubsection{The pattern of increasing organism diversity}
Qualitatively we observe that the number of species generated are monotonically increasing (Fig.~\ref{fig:organism_diversity}), albeit at a decreasing rate.
The ratio between the number of unique species and the generation index is 2:1 at index=200, and approaches 4:1 near index=800.
This suggests more mechanisms are generated and become concentrated on individual species (\eg for grasshoppers, there might be several distinct mechanisms relevant to the problem of `managing turbulence' such as their foldable wing structures, lightweight exoskeleton designs, or joints in their legs enabling repeated high jumps) on average as generation continues. 
In our future work, we will explore whether and how the mechanisms generated for the same species semantically differ from one another.
We also plan to examine how our generation approach compares to other baseline approaches in terms of the efficiency of generating mechanisms across diverse organisms, by measuring the slope of organism diversity over generation index. 

\subsection{Case Studies}
\subsubsection{`Design a secure bike rack for sedans'} \label{appendix:case_study_1}
\begin{figure}[h]
    \includegraphics[width=\linewidth]{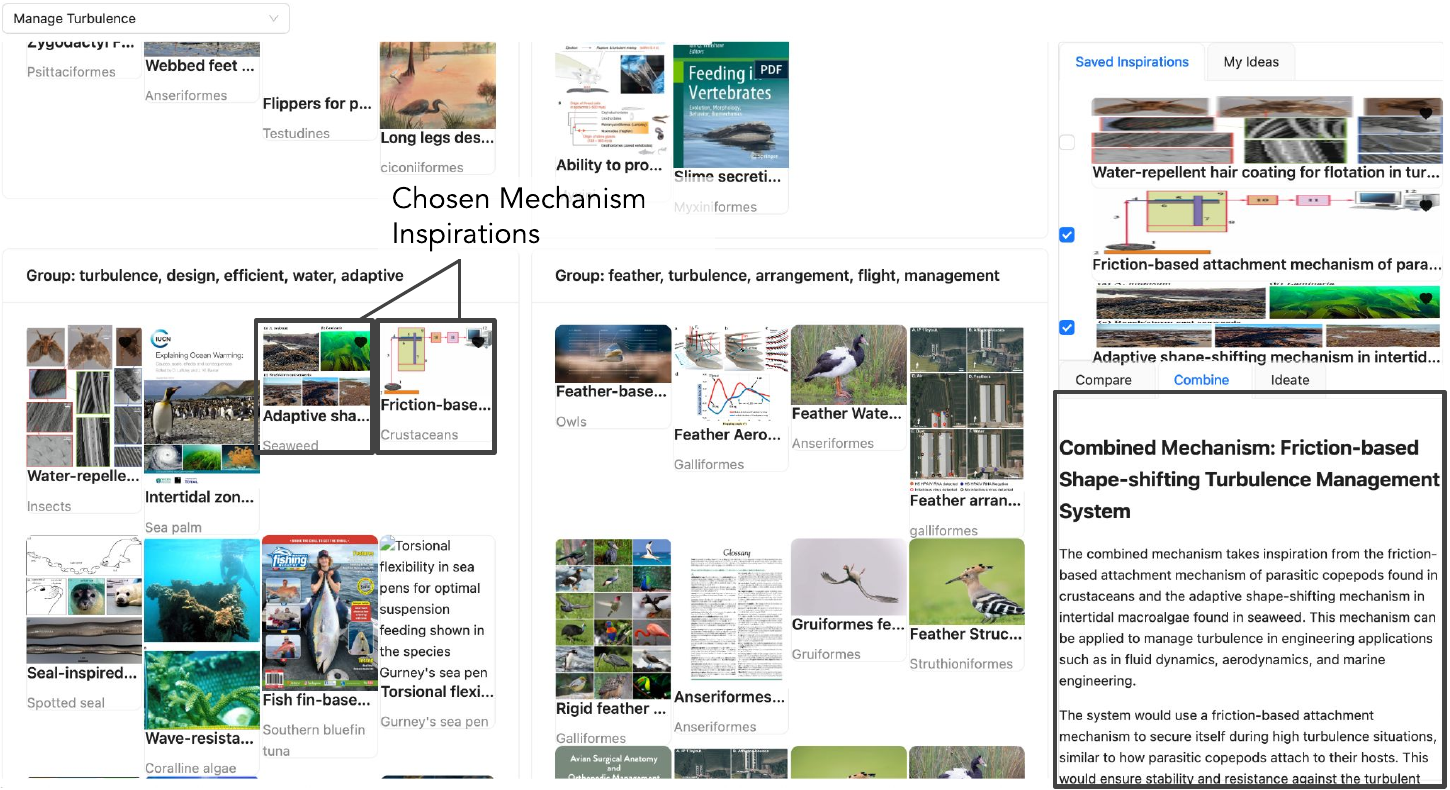}
    \vspace{-1.5em}
    \caption{\sys{} interface featuring biological mechanisms for the query design challenge `Manage Turbulence'. User-chosen mechanisms are: ``Adapative shape-shifting mechanism in intertidal microalgae'' and ``Friction-based attachment mechanism of parasitic copecods.'' A new combination idea is shown in the `Combine' tab, facilitating further synthesis and ideation of a new idea.}
    \vspace{-1em}
    \label{fig:case_study_1}
\end{figure}
Fig.~\ref{fig:case_study_1} shows user interactions based on the motivating usage scenario described in \S\ref{section:usage_scenario}.
An accompanying demonstration video can also be accessed at: \url{https://drive.google.com/file/d/1qD1rJXsE8g1NimLnd2quF_ofR6KFXdrh/view?usp=drive_link}.

\subsubsection{`Design robust collision resistance for racing cars'}
The details of this case study can be found in Appendix \S\ref{appendix:case_study_2}.
\section{Discussion and Future Work} \label{appendix:future_work}
\subsection{Closing the `Gulf of Geometric Transfer' with combinatoric visual mechanism generation}
\begin{figure}[h]
    \includegraphics[width=\linewidth]{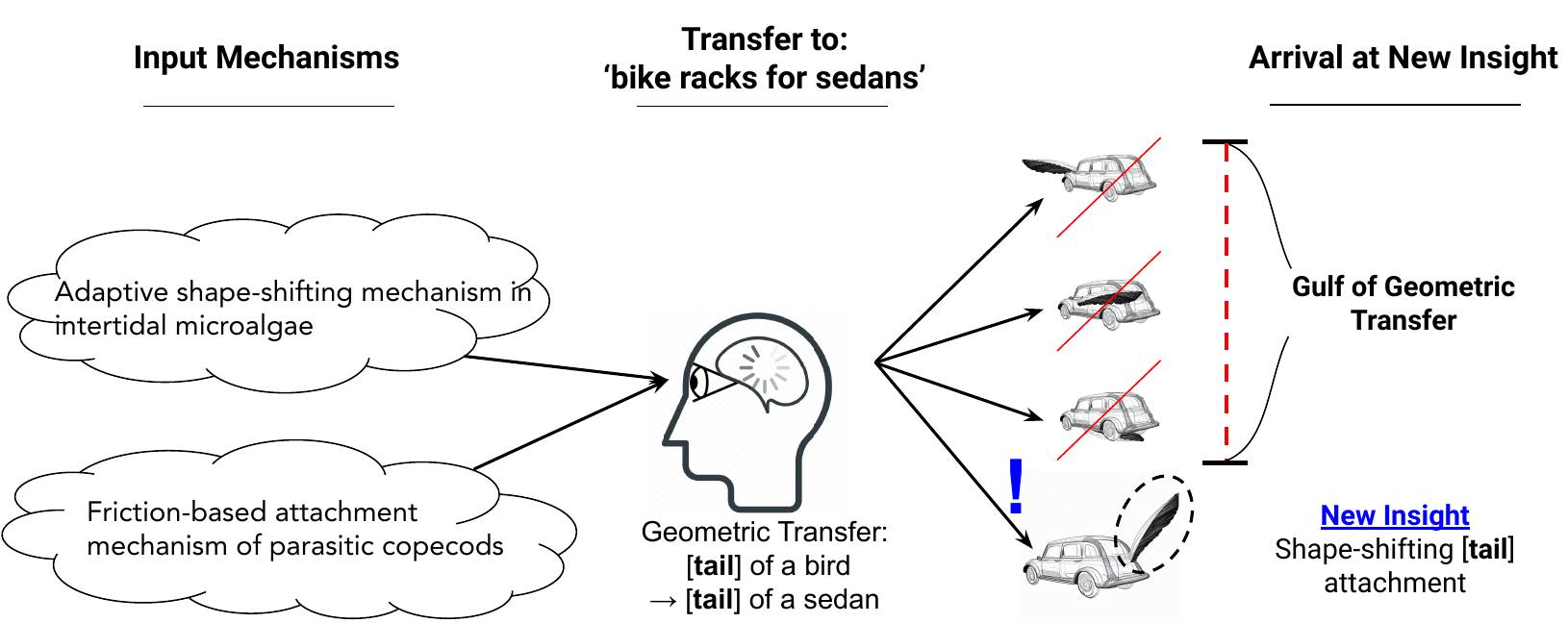}
    \vspace{-1.5em}
    \caption{When the designer arrives at the new insight of designing a tail-like, shape-shifting bike rack attachment (Appendix~\ref{appendix:case_study_1}), she is combining two source mechanisms and transferring the new idea into a new geometry in the target domain (`at the \emph{tail} of a sedan') that could suit the combined mechanism. The successful generation of this combination comes after a series of failed attempts to generate geometries, which we call the \textbf{Gulf of Geometric Transfer}.}
    \label{fig:future_work_closing_gulf_of_transfer}
\end{figure}
The designer in Case Study 1 (Appendix~\ref{appendix:case_study_1}) arrives at a successful transfer of mechanisms naturally after several unsuccessful iterations of generation.
A key process in transferring a new idea into the target domain involves visualizing a candidate geometry, such as in this case `a bike rack attachment to the \emph{tail} of a sedan'.
The process of identifying an effective geometry for transfer like `at the tail of a sedan' can resemble a combinatoric search, where the designer visualizes the location of integration for the new idea in the target object, from a possible source object and its geometry, such as `the tail of a bird'.
The process of evaluating an effective geometry for transfer may further involve analogical processing, for instance, by comparing the integrated geometry to familiar objects and judging the aesthetic and functional value of such a geometry.

We hypothesize that one challenge of this transfer process is what we term `\textbf{Gulf of Geometric Transfer}' (see Fig.~\ref{fig:future_work_closing_gulf_of_transfer}, right; the concepts related to this namesake in design and usability studies are the Gulf of Execution and the Gulf of Evaluation~\cite{gulf_of_execution_gulf_of_evaluation}), which refers to the failed iterations that one has to endure before obtaining a successful geometric insight that supports the new idea.
To bridge this gulf, we envision a future work area that aids designers in visualizing what successful geometries that facilitate transfer of ideas might look like, and also helps them eliminate any blind spots in the process that could hinder new insights.
A fruitful starting point to this end may involve generating visual representations of geometries that are related to the new abstract idea from various key orientations, such as \{\code{up}, \code{down}, \code{front}, \code{rear}, \code{left}, \code{right}, \code{internal}, \code{external}\}.
Enumerating these geometries visually could assist designers in effectively reviewing possible geometries and facilitating their imagination for alternative geometries that may be useful for further consideration.

\subsection{Supporting Generative `Frame Slips'}
The geometric transfer discussed in the previous section can be understood as one form of a more generalized phenomenon in analogical transfer, where a conceptual frame `slips' into a different one, thereby generating novel insight.
For example, Niels Bohr transferred the concept of planets orbiting the sun to the domain of atomic structure to propose a new model in which electrons orbit the nucleus in a similar manner~\cite{gentner1983structure}.
This new structure of the atom significantly differed from the dominant `plum-pudding' model of the time, leading to a \emph{re-structuring} of the concept~\cite{gentner2002analogy}.
Here, \emph{re-structuring} is related to conceptual frames that successfully `slip' into a new, analogous problem domain, causing structural changes in the conceptaul framework of the receiving domain or vice versa.
In the case studies, the concept of the `tail of a car' slipped into geometrically and syntactically similar `tail of a bird', engendering a new insight.
Similarly while studying how crowds generate design concepts for children's chairs, Yu et al.~\cite{swing_slide_chairs} found that the concept of a playground `slide' slipped into another semantically related playground concept `swing set', which then gave a novel design idea of swing chairs.

Understanding the process of `\textbf{Frame Slipping}' is therefore crucial for improving analogical transfer.
Existing examples of `frame slips' imply that such transitions can occur through different relational paths, for example, syntactic, semantic, functional, abstracted commonality, geometric, or other means.
While geometric transfer may impose a more concrete form of transformation onto the transfer process, other means of transfer may be less understood, possibly involving more complex multimodal processes that combine multiple relational paths at once.
Nonetheless, significant value of an inspiration often results from enabling frame slips to new parts of the design space, opening up not only a single point of an idea, but an entire space of ideas for future exploration.

\subsection{Supporting any problem queries by incorporating rich problem-mechanism relations}
One of \sys's limitations is its fixed problem queries.
Though the five pre-generated problem queries provide a useful entry to mechanism organisms that may be applicable to a diverse set of design challenges, it comes at the cost of an inability to query biological mechanisms for \emph{any} engineering design problems described in natural language text.
As designers and engineers progress in interacting with the system, they may naturally come up with follow-up queries that may differ from the source queries, that could emphasize important constraints around the design problem, or specify low-level details newly understood to be important to consider.
Adaptation to such evolving user query intent requires further personalization and scaffolding in the workflow.
In future work, mixed-initiative workflows may leverage user interaction traces as input to LLM operations (\cf~\cite{KANG2023Synergi,liu2023selenite}) to augment query input and automatically search data to retrieve analogical results.

In order to enable search by free-text problem queries, the underlying data model needs to be extended to contain multiple problem-mechanism relations beyond the single problem present in the schema $\{\forall i|(p_i, m_i, o_i)\}$, and into an enriched dataset with mappings between problems $p_1, p_2, \cdots, p_n \in P$ and an applicable mechanism $m_k$, as commonly the case in engineering (\eg `spider silk' can be used for multiple engineering challenges such as replacing steel bars in concrete or wound suture and prosthesis~\cite{gu2020mechanical}).

One way to expand the rich problem-mechanism relations in a scalable manner is to prompt LLMs to come up possible engineering design problems that a given biological mechanism could be applied to.
Here, na\"{i}ve prompting may suffer from conceptual redundancy, analogous to the challenge of curating diverse mechanisms, that limits the diversity in mechanism-to-problem mappings.
Another approach may be to intelligently use the existing dataset $\{\forall i|(p_i, m_i, o_i)\}$ to identify similar mechanisms $m_i$ in $(p_i, m_i, o_i)$ and $m_j$ in $(p_j, m_j, o_j)$ that can be mapped onto disparate problems: $m_i \sim m_j \rightarrow p_i, p_j$.
This approach however assumes the presennce of many correlated such mechanisms with disparate problem pairs in the dataset, which need empirical examination for support. Once the enriched dataset $\{\forall i|m_i \rightarrow S(m_i)\}$ (Here, $S(m_i)$ denotes the set of engineering design problems that $m_i$ is applicable to) is made available, one simple approach for allowing querying on any problem text is to construct a similarity search index (\eg the HNSW index of FAISS~\cite{johnson2019billion}) using a chosen text embedding approach.

\subsection{Supporting improved visual representation of biological mechanisms}
Future work may focus on promising methodologies for improved visual representation of biological mechanisms.
Augmentative approaches to the simple retrieval approach described in the current prototype include `Retrieve and Filter' and `Highlight Species Subparts'.
The `Retrieve and Filter' method involves a process of purging non-representative and low-quality images and ranking the remaining ones based on their semantic distance to the mechanism text, resulting in images with higher semantic coherence with the corresponding mechanism description.
On the other hand, the `Highlight Species Subparts' method will involve sourcing for representative species illustrations or photographs, and highlighting components corresponding to mechanisms via a translucent overlay.
This method can be applied when previous method fails to yield high-quality images.

While `Highlight Species Subparts' presents potential, we also anticipate certain limitations with it.
For example, specific mechanisms could only apply to a tiny animal area, rendering a whole body image inappropriate, such as in cases of the adhesive paws of geckos or the hierarchical layers in pomelo peels.
Additionally, in certain cases, embedding textual labels within a diagrammatic representation, as opposed to using realistic species images, could provide more beneficial insights, annotating key components rather than merely a visual highlight.
Understanding the trade-off between these strategies could be beneficial.

Alternatively, generative AI could unlock new approaches to generate representative mechanism images on demand.
Here, a systematic method such as focusing on cases where retrieval-based images fail could prove effective.
A practical approach would involve creating a mechanism taxonomy potentially in tabular format showcasing a mechanism's text, its type, and specific example images that unsuccessful retrieval-based approaches attempted to represent.
This could offer valuable insights into the areas and reasons why our retrieval-based imaging models are not successful, thereby guiding the further development of generatively-focused models.
\section{Conclusion}
In this work we present \sys, an end-to-end system for generating a biological-analogical mechanisms dataset and an interactive interface that facilitates learning new biological mechanisms for a design challenge, and synthesizing new solution ideas inspired by analogical mechanisms.
\appendix
\section{Taxonomy Generation} 
\subsection{Prompt Used for Taxonomy Generation}
\label{appendix:taxonomy_prompt}
The prompt used for taxonomy generation for LLMs can be found in Fig.~\ref{fig:taxonomy_prompt}.
\begin{figure*}[htbp]
\noindent\rule{\textwidth}{0.4pt}
\begin{lstlisting}
[System Message]
You are an expert biologist who knows species and their taxonomic hierarchy very well. Follow the instructions to the letter.
- Return the scientific term for each taxonomic rank the species belongs to.
- Enclose keys and values using double quotes ("...") and format them into a Python dictionary.
- Use the taxonomic ranks as keys and corresponding scientific terms as their values.
- Do not add any other text.
\end{lstlisting}
\noindent\rule{\textwidth}{0.4pt}
\begin{lstlisting}
[User Message]
What {"domain", "kingdom", "phylum", "class", "order", "family", "genus"} does "{organism}" belong to? Format your reply into a Python dictionary.
\end{lstlisting}
\noindent\rule{\textwidth}{0.4pt}
\vspace{-1.5em}
\caption{The prompt used to generate the taxonomy of each organism.}
\label{fig:taxonomy_prompt}
\end{figure*}

\subsection{Complete List of Organisms Used for Taxonomy Generation} \label{appendix:complete_list_of_organisms}
\code{\{`spider monkey', `prairie dog', `garden tiger moth', `african sacred ibis', `argiope argentata', `ostrich', `groundhog', `danio rerio', `gannet', `deer', `cattle', `glyptodon', `alligator snapping turtle', `leopard', `arctic ground squirrel', `cormorants and shags', `bears', `squirrels', `herons', `european badger', `golden silk orb-weaver', `aardvark', `seahorses', `banner-tailed kangaroo rat', `hyenas', `pink fairy armadillo', `giant otter', `bighorn sheep', `hippopotamus', `california ground squirrel', `european bee-eater', `beech marten', `leopard gecko', `tailorbird', `testudinidae', `emperor penguin', `northern pike', `giant clam', `stoat', `horse', `nutria', `tree-kangaroo', `giraffe', `guinea baboon', `ferret', `bonytail chub', `baya weaver', `brook trout', `pelican', `mallard', `roseate spoonbill', `mountain weasel', `pocket gophers', `lybia edmondsoni', `giant anteater', `common raccoon dog', `dewdrop spiders', `armadillo girdled lizard', `arctic fox', `bison', `swordfish', `bald eagle', `chimpanzee', `asbolus verrucosus', `sperm whale', `abalone', `golden jackal', `hornet', `zebra', `orangutans', `peregrine falcon', `atlantic cod', `burrowing owl', `african wild dog', `maned wolf', `honey bee', `naked mole-rat', `echidnas', `bowerbirds', `rhinoceros', `beaver', `bombyx mori', `common box turtle', `hummingbird', `domestic sheep', `wolverine', `raccoon', `evergreen bagworm', `pig', `muskrat'\}}

\subsection{Error analysis} \label{appendix:taxonomy_generation_error_analysis}
We find that some error cases in taxonomy generation could be attributed to recent changes in classification in the literature.
For example,  both GPT4 and GPT3.5-turbo models classified naked mole-rats as then literature-accepted `Bathyergidae' for their family, same as other African mole-rats.
However, more recently naked mole-rats were placed in a separate family, Heterocephalidae~\cite{Naked_mole_rat}.

Among the error cases overlapping between the two models, we found cases that either the GPT3.5-turbo or the GPT4 model wins over the other (\eg for `hummingbird', GPT3.5-turbo generated `archilochus' as its genus whereas GPT4 generated `various'; for `boxer crab', GPT3.5-turbo generated `hymenoptera' which is an order of insects, whereas GPT4 generated `decapoda', which is the correct order).
In other cases, both models outputted similarly incorrect answers, for example for `sea snail', GPT3.5-turbo generated `neogastropoda' whereas GPT4 generated `archaeogastropoda' (the Wikipedia gold answer was `lepetellida').

\section{Additional Case Study: `\textit{Design robust collision resistance for racing cars}'} \label{appendix:case_study_2}
\begin{figure}[h]
    \includegraphics[width=\linewidth]{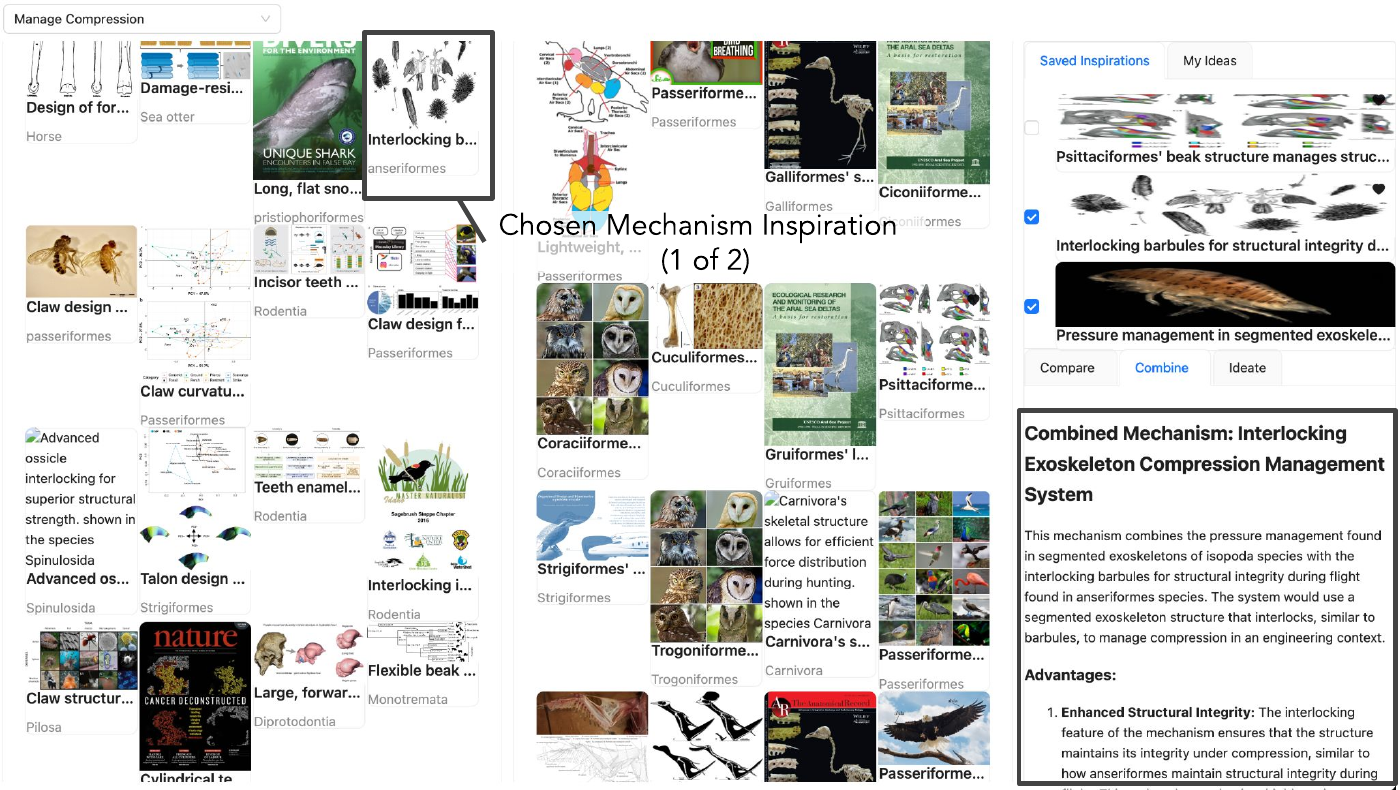}
    \vspace{-1.5em}
    \caption{\sys{} displays biological mechanisms for the query design challenge `Manage Compression'. User-chosen mechanisms are: ``Interlocking barbules for structural integrity during flight'' and ``Pressure management in segmented skeletons.''}
    \label{fig:case_study_2}
\end{figure}
For another designer working on improving collision resistance for racing cars, the core challenge might be the management of compression during a collision.
When the designer clicks on the `Manage Collision' problem as a query (Fig.~\ref{fig:case_study_2}, top right), \sys{} once again displays biological mechanism clusters related to the problem.
He finds ``Interlocking barbules for structural integrity during flight'' and ``Pressure management in segmented skeletons,'' interesting mechanisms employing different approaches to compression management (the former commonly found in feathers, the latter among isopods).
He selects the `Combine' tab to obtain design ideas that could benefit from both mechanisms.
The result describes an approach that applies interlocking structures to exoskeleton segments, providing him with insight into a `sliding-lock' mechanism for modular panels of the racing car body.
This mechanism allows for extensive compression upon collision, thus increasing impact time (subsequently decreasing peak impact force) and more effectively dissipating force to non-critical body components. A video demonstrating user interaction can be also accessed at: \url{https://drive.google.com/file/d/1F2z1AX2aTMlpAzNX9kyePuMebFsB0-Wx/view?usp=sharing}.

\begin{ack}
This research was supported by the Toyota Research Institute.
\end{ack}

\bibliographystyle{plain}
\bibliography{neurips_2023}








\end{document}